\documentstyle[aps,preprint]{revtex}
\def\pr#1#2#3{{\it {Phys. Rev.}} {\bf{D#1}}, #2 (#3)}
\begin{document}

\preprint{9803339}
\title{Radiative Breaking of  Gauge Symmetries in the MSSM and in 
its Extensions}
\author{Athanasios Dedes{\footnote{E-mail: adedes@cc.uoi.gr}}}
\address{Division of Theoretical Physics, Physics Department, University of 
Ioannina, GR-45110, Greece}

\maketitle

\vspace*{0.4in}

\centerline{A Dissertation in Physics{\footnote{The
full postscript version in Greek (171 pages) is available from \newline \indent
\indent
{\tt http://artemis.sci.uoi.gr/\~{}adedes/phd.ps.gz}}}}
\centerline{Dissertation Supervisor: Prof. K. Tamvakis}

\begin{abstract}
The radiative electroweak symmetry 
breaking, the $b-\tau$  Yukawa and gauge couplings unification
in the MSSM and its $SU(5)$ extensions are studied in detail.
In the framework of the two-loop renormalization group equations
both low- and high-energy threshold effects are included.
In the case of the minimal $SU(5)$, the
values of $\alpha_s$ obtained are somewhat larger than 
the experimental average. The Peccei-Quinn version of the
missing-doublet $SU(5)$ model generally predicts smaller
values of $\alpha_s$ and
$b-\tau$ unification, in excellent agreement with all
low energy experimental data.
We also study
 the generation of the GUT scale through radiative corrections
in the context of an $R$-symmetric ``flipped" $SU(5)\times U(1)_X$ model
and we  find that this is possible in a certain range of values
of the parameter space.

\end{abstract}

\newpage

\centerline{\bf Summary of the Dissertation}
In this thesis, we  study 
 Radiative  Symmetry Breaking in the Minimal Supersymmetric
 Standard Model (MSSM) and in extensions of it. The current theoretical
and experimental  situation for the Standard Model (SM) and the MSSM is
reviewed and the solution of the hierarchy problem in the case
of the MSSM is given in detail. The realization of the dimensional
transmutation mechanism in the MSSM is also treated in the same manner.
Effective field theoretical techniques and  the concept of 
thresholds  are presented.
We make a complete analysis of radiative electroweak symmetry 
breaking in the MSSM and its $SU(5)$ extensions
including low- and high-energy threshold effects
in the framework of the two-loop renormalization group.
This program
requires the calculation of all wavefunction, vertex  and mass
renormalizations for all particles involved.
We compute the mass spectrum of sparticles
and Higgses consistent with the existing experimental constraints.
In the MSSM we find that 
the effect of the threshold corrections is in general of the same order of
magnitude  as the two-loop contributions with the exception of the
heavy Higgses and those neutralino and chargino states that are nearly
Higgsinos for large values of the parameter $\mu$.
We have also considered {\it minimal} $SU(5)$, the
{\it missing-doublet} $SU(5)$, a
 {\it Peccei-Quinn} invariant version of
$SU(5)$. 
We derive permitted ranges
for the parameters of these models in relation
to predicted  $\alpha_{s}$ and $M_G$
values within the present experimental accuracy.
The parameter regions allowed under the constraints
of radiative symmetry breaking, perturbativity and proton
stability, include the experimentally designated domain for
$\alpha_s$. In the case of the {\it minimal} $SU(5)$, the
values of $\alpha_s$ obtained are somewhat large in
comparison with the experimental average. The
{\it missing-doublet} $SU(5)$, generally, predicts smaller
values of $\alpha_s$. In both versions of the {\it missing-doublet},
the high energy threshold effects on $\alpha_s$ operate in the
opposite direction than in the case of the minimal model, leading
to small values. In the case of the {\it Peccei-Quinn} version
however the presence of an extra intermediate scale allows us to
achieve an excellent agreement with the experimental $\alpha_s$
values. 
We have also made  a complete analysis of the Yukawa coupling unification
in $SU(5)$ extensions of the MSSM in the framework of the radiative
symmetry breaking scenario.
Both logarithmic and finite threshold
corrections of sparticles have been included in
the determination of the gauge and Yukawa couplings at $M_Z$. The
effect of the heavy masses of each model in the renormalization
group equations is also included. We find that in the minimal
$SU(5)$ model $b-\tau$ Yukawa unification can be achieved
for too large a value of $\alpha_s$.
On the other hand the {\it
Peccei-Quinn} version of the {\it Missing Doublet} model,
with the effect of the right handed neutrino also included,
exhibits $b-\tau$ unification in excellent agreement with all
low energy experimental data. Unification of all Yukawa couplings
is also discussed.
Finally, we have studied
 the generation of the GUT scale through radiative corrections
in the context of a $R$-symmetric ``flipped" $SU(5)\times U(1)_X$ model.
A negative mass squared term for the GUT Higgs fields
develops due to radiative effects along a flat direction
at a superheavy energy scale. The $R$-symmetry is essential in
maintaining triplet-doublet splitting and $F$-flatness in the
presence of non-renormalizable terms. The model displays radiative electroweak
symmetry breaking and satisfies all relevant phenomenological constraints.

\vspace*{1in}

\centerline{\bf List of Publications}

\vspace*{1mm}

[1] A. Dedes, A. B. Lahanas and K. Tamvakis, \pr{53}{3793}{1996}.

[2] A. Dedes, A. B. Lahanas, J. Rizos and K. Tamvakis, \pr{55}{2955}{1997}.

[3] A. Dedes and K. Tamvakis, \pr{56}{1496}{1997}.

[4] A. Dedes, C. Panagiotakopoulos and K. Tamvakis, hep-ph/9710563,\newline
\indent \indent
{\it to appear in Phys. Rev. D}.

\newpage

\centerline{\bf Contents}
{\bf 1. Introduction \dotfill 1} \\[1mm]
\indent  1.1 Introduction to Supersymmetry \dotfill 10 \\ 
\indent 1.2 Construction of Supersymmetric Field Theories \dotfill 10 \\
\indent \indent  1.2.1 The Supersymmetric algebra\dotfill 11 \\
\indent \indent  1.2.2  Chiral and Vector Superfields \dotfill 13 \\
\indent \indent  1.2.3  Construction of the N=1 Lagrangian
\dotfill 16 \\
\indent  1.3 Quadratic Divergences \dotfill 19 \\
\indent  1.4 Supersymmetry Breaking \dotfill 25 \\[1mm]
{\bf 2.   The Minimal Supersymmetric Standard Model (MSSM)
\dotfill 29} \\[1mm]
\indent  2.1 Introduction
\dotfill 29 \\
\indent  2.2 Definition of the MSSM \dotfill 29  \\
\indent  2.3 Radiative Electroweak Symmetry Breaking in the MSSM 
\dotfill 39 \\
\indent \indent 2.3.1  Dimensional Transmutation
 \dotfill 39 \\
\indent \indent 2.3.2 
 Realization of the dimensional transmutation mechanism in the MSSM
\dotfill 41 \\
\indent  2.4 The Spectrum of the MSSM \dotfill 46 \\[1mm]
{\bf 3.  Effective Field Theory Technics and  Low energy data 
\dotfill 53} \\[1mm]
\indent  3.1 Introduction
\dotfill  53 \\
\indent  3.2  Weak interactions at low energies : Tree level
\dotfill 53 \\
\indent  3.3 Renormalization in effective field theories 
\dotfill 55 \\
\indent   3.4  Decoupling of heavy particles \dotfill 58 \\
\indent   \indent 3.4.1  
Mass dependent scheme 
\dotfill 58 \\
\indent   \indent 3.4.2  
$\overline{MS}$ and $\overline{DR}$
 mass independent schemes
\dotfill 59 \\
\indent  3.5 The electroweak mixing angle \dotfill 61 \\
\indent  3.6 The strong coupling constant \dotfill 64 \\
\indent  3.7 Experimental constraints on Supersymmetric and Higgs particles 
\dotfill 66
 \\[5mm]
{\bf 4.  Radiative Electroweak Symmetry Breaking in the MSSM and Low Energy
\newline \indent Thresholds
\dotfill 69} \\[1mm]
\indent  4.1 Introduction \dotfill  69 \\
\indent  4.2 The Renormalization Group and threshold effects
\dotfill 69\\
\indent  4.3 Formulation of the problem and numerical analysis \dotfill 75\\
\indent  4.4 Conclusions \dotfill 78 \\[1mm]
{\bf 5. Threshold Effects and Radiative Electroweak Symmetry Breaking
\newline \indent in 
SU(5) Extensions of the MSSM
\dotfill 91} \\[1mm]
\indent  5.1 Introduction 
\dotfill 91 \\
\indent  5.2 Minimal SU(5) \dotfill 94 \\
\indent  5.3 Missing Doublet Model 
\dotfill 98 \\
\indent  5.4 Peccei-Quinn symmetric Missing Doublet Model
\dotfill 102 \\ 
\indent  5.5 A version of SU(5) with light remnants 
\dotfill 106 \\
\indent  5.6 Conclusions \dotfill 110 \\[1mm]
{\bf 6. $b$-$\tau$ Unification in SU(5) Extensions of the MSSM 
\dotfill 111}\\[1mm]
\indent  6.1 Introduction \dotfill 111\\
\indent  6.2  $b$-$\tau$ Unification in the Minimal SU(5) Model
\dotfill 113\\
\indent  6.3  $b$-$\tau$ Unification in the Peccei-Quinn version of the
Missing Doublet Model
\dotfill 114\\
\indent  6.4 Conclusions \dotfill 122 \\[1mm]
{\bf 7. Radiative GUT Symmetry Breaking in an R-Symmetric
\newline \indent Flipped SU(5)
Model
\dotfill 125}\\[1mm]
\indent  7.1 Introduction \dotfill 125\\
\indent  7.2 An R-Symmetric version of $SU(5)\times U(1)_X$
\dotfill 126\\
\indent  7.3 Radiative Corrections and the RGEs \dotfill 128 \\
\indent  7.4 Numerical Analysis and Conclusions \dotfill 132 \\[1mm]
{\bf 8. Conclusions and Outlook \dotfill 135} \\[1mm]
{\bf Appendix A:}  Calculations in the Superspace \dotfill 139\\[2mm]
{\bf Appendix B:}  Two loop RGEs in the MSSM 
\dotfill 143 \\[2mm]
{\bf Appendix C:} Numerical Calculations  \dotfill 151 \\[2mm]
{\bf Appendix D:} RGEs above $M_G$ in the Minimal SU(5) 
\dotfill 153\\[2mm]
{\bf Appendix E:}    Description of the SU(5) Models   
\dotfill 155 \\[2mm]
{\bf Bibliography} \dotfill 163 

\end{document}